\pdfoutput=1
\documentclass[Physsubmission, Phys]{SciPost}

\hypersetup{
    colorlinks,
    linkcolor={red!50!black},
    citecolor={blue!50!black},
    urlcolor={blue!80!black}
}

\usepackage[bitstream-charter]{mathdesign}
\DeclareSymbolFont{usualmathcal}{OMS}{cmsy}{m}{n}
\DeclareSymbolFontAlphabet{\mathcal}{usualmathcal}

\begin{document}

\begin{center}{\Large \textbf{
      Amplitude and colour evolution
}}\end{center}

\begin{center}
Simon Pl\"atzer\textsuperscript{1,2,3$\star$}
\end{center}

\begin{center}
{\bf 1} Institute of Physics, NAWI Graz, University of Graz,\\Universit\"atsplatz 5, A-8010 Graz, Austria
\\
{\bf 2} Particle Physics, Faculty of Physics, University of Vienna,\\Boltzmanngasse 5, A-1090 Wien, Austria
\\
{\bf 3} Erwin Schr\"odinger Institute for Mathematics and Physics,\\University of Vienna, A-1090 Wien, Austria
\\
* simon.plaetzer@uni-graz.at
\end{center}

\begin{center}
\today
\end{center}

\definecolor{palegray}{gray}{0.95}
\begin{center}
\colorbox{palegray}{
  \begin{tabular}{rr}
  \begin{minipage}{0.1\textwidth}
    \includegraphics[width=23mm]{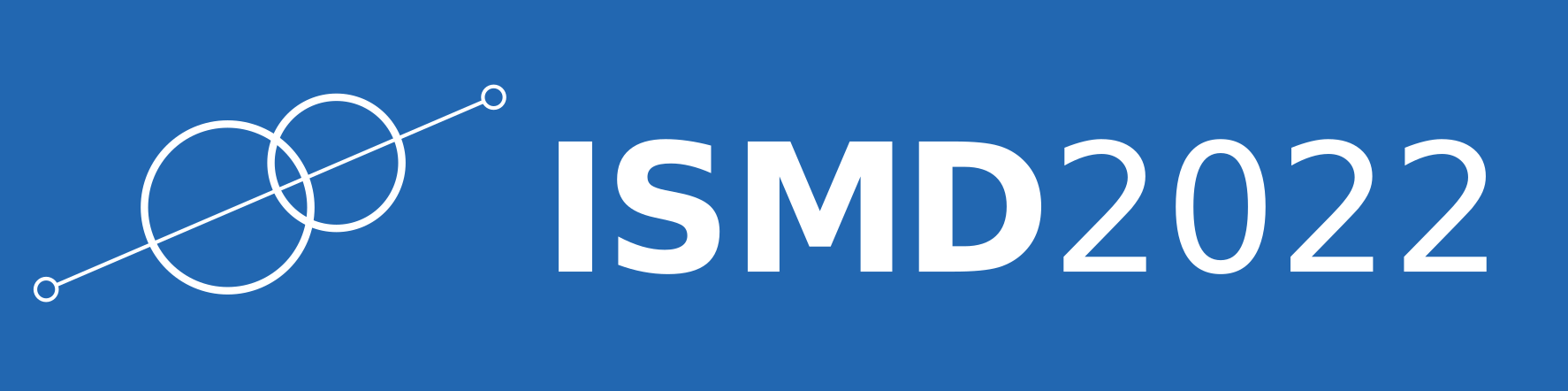}
  \end{minipage}
  &
  \begin{minipage}{0.8\textwidth}
    \begin{center}
    {\it 51st International Symposium on Multiparticle Dynamics (ISMD2022)}\\ 
    {\it Pitlochry, Scottish Highlands, 1-5 August 2022} \\
    \doi{10.21468/SciPostPhysProc.?}\\
    \end{center}
  \end{minipage}
\end{tabular}
}
\end{center}

\section*{Abstract}
         {\bf Colour evolution and parton branching at the amplitude
           level have become important theoretical frameworks to
           improve parton showers, and are algorithms in their own
           right: they complement shower development by resummation
           algorithms capable of including interference effects and
           subleading colour contributions at an unprecedented
           level. I summarize recent development in the field,
           focusing on soft gluon evolution, hadronization, and the
           CVolver framework.}

         \vspace{10pt}
\noindent\rule{\textwidth}{1pt}
\tableofcontents\thispagestyle{fancy}
\noindent\rule{\textwidth}{1pt}
\vspace{10pt}

\section{Introduction}
\label{sec:intro}

Event generators are central to phenomenology at colliders. Parton
showers are the central component of event generators which describe
the build-up of jets through energy loss of primary partons with large
momenta of the order of some hard scale $Q$ down to typical scales of
$\mu_S \sim 1 {\rm GeV}$, where phenomenological models of
hadronization describe how a partonic ensemble is forming the observed
hadronic final state.

The accuracy of parton shower algorithms can only be assessed for
certain (classes of) observables. Coherent branching algorithms such
as the one underpinning the Herwig event generator
\cite{Gieseke:2003rz} are known to predict global observables at
next-to-leading logarithmic (NLL) accuracy, and dipole-type showers
{\it e.g.} the Herwig implementations following
\cite{Platzer:2009jq,Platzer:2011bc}, predict non-global observables
at leading logarithmic (LL) accuracy. While all of these algorithms
are based on the large-$N_c$ limit (with $N_c=3$ the number of colours
in QCD), the structure of coherent branching is actually able to
account for subleading colour provided that the structure of the hard
process is sufficiently simple \cite{Forshaw:2021mtj}. Dipole showers
have recently been improved to reproduce properties of coherent
branching and are thus now able to also predict global observables at
the NLL level, work which had been initiated in
\cite{Dasgupta:2020fwr,Forshaw:2020wrq}. With an additional
algorithmic tweak \cite{Hamilton:2020rcu,Holguin:2020joq} they can
even reproduce the full-$N_c$ colour factors as dictated by QCD
coherence.

Several of the above mentioned refinements are based on approaches
which analyse properties of amplitudes as they build up through
successive radiation: Parton branching at the amplitude level
\cite{Nagy:2014mqa,Forshaw:2019ver} originates from the study of
multiple soft gluon emission
\cite{Platzer:2013fha,AngelesMartinez:2018cfz} and has become both a
theoretical method as well as an algorithm in its own right which
implements evolution at the amplitude level, or more precisely at the
level of the cross section density operator which we shall introduce
later. In this contribution, we will focus on how these algorithms can
be constructed (at least in the soft gluon case), how we can use such
constructions to assess the accuracy at which the evolution can
predict certain observables, and how hadronization models could appear
and be constrained within parton branching at the amplitude level.

\section{Cross sections and cross section density operators}
\label{sec:densityoperators}

\begin{figure}
  \begin{center}
  \includegraphics[scale=0.3]{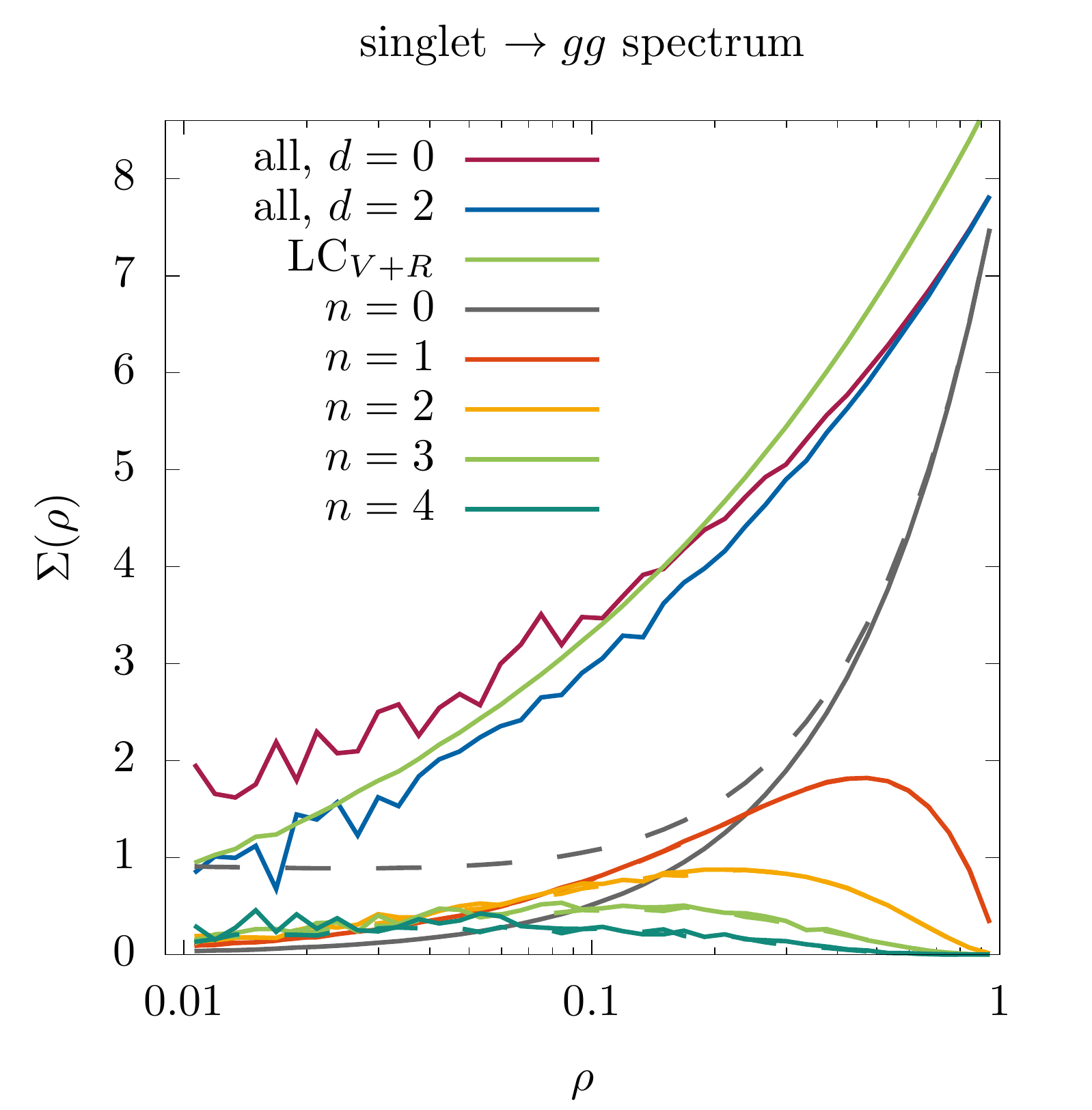}
  \includegraphics[scale=0.3]{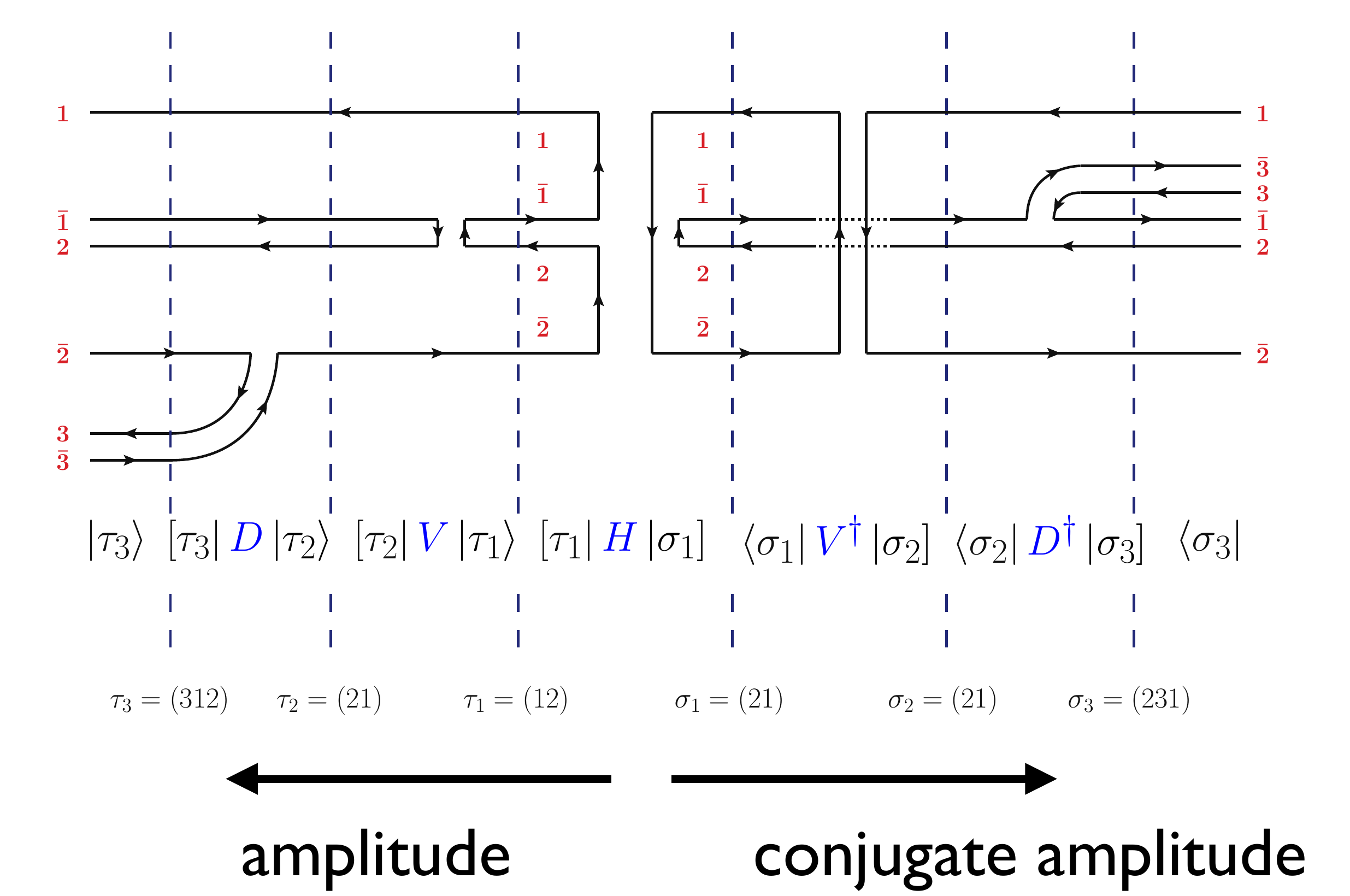}
  \end{center}
  \caption{\label{fig:ampevolution}Example results from amplitude
    evolution in the colour flow basis. We calculate the
    gaps-between-jets cross section with a veto scale $\rho$ for
    cone-like jets. The evolution proceeds by a Monte Carlo over
    colour flows resulting from gluon exchange and emission in the
    amplitude and conjugate amplitude, allowing all histories to
    interfere with each other. Here $V$ denotes the insertion of the
    virtual evolution operator resulting from the exponentiation of
    ${\mathbf \Gamma}$ and $H$ is the initial condition provided by
    the amplitude of the hard scattering process. Figures amended from
    \cite{DeAngelis:2020rvq}.}
\end{figure}

The aim of parton branching at the amplitude level is to calculate
cross sections for a hard process accompanied by an additional number
of partons which are either soft (with momentum components $\ll Q$),
or collinear by building up the according scattering amplitude and
squaring it at the end of the evolution. This is contrary to standard
parton showers, which would iterate additional emissions in a
probabilistic manner so long as they factorise at the level of the
cross section for each individual additional radiation. Amplitude
evolution is required whenever we cannot constrain the kinematics of
additional radiation in a way that coherence arguments would allow to
simplify the cross section. Colour correlations then unavoidably
persist and can only be simplified further if one resorts to the
large-$N_c$ limit. The large-$N_c$ limit itself is problematic since
in a parametric counting for a logarithmic enhancement of $\alpha_S
L^2\sim 1$ (with $L$ a large logarithm of the observable quantity),
subleading logarithmic corrections might appear at the same level as
subleading colour corrections since also $\alpha_s N_c^2\sim
1$. Effects due to the exchange of gluons in the Glauber region can
also not be accounted for in the large-$N_c$ limit.  Amplitude
evolution algorithms rest on a decomposition of amplitudes in colour
space,
\begin{equation}
  |{\cal M}\rangle = \sum_\sigma {\cal M}_\sigma |\sigma\rangle
\end{equation}
where the abstract vectors $|\sigma\rangle$ (not to be confused with a
true quantum mechanical state) span a space of colour structures, {\it
  i.e.} tensors of ${\rm SU}(N)$. In general, other quantum numbers as
well as spin might be considered in a similar manner and we have
recently been pointing out how such a formalism would be carried over
to the entire Standard Model \cite{Platzer:2022nfu}. For $n$
additional emissions on top of a hard process, amplitude evolution
then considers the cross section density operator ${\mathbf A}_n$,
which relates to the amplitude $|{\cal M}_n\rangle$ and the cross
section with $n$ additional partons as
\begin{equation}
{\mathbf A}_n = |{\cal M}_n\rangle \langle {\cal M}_n| \ , \qquad {\rm d}\sigma_n = {\rm Tr}\left[{\mathbf A}_n\right] {\rm d}\phi_n \ .
\end{equation}
The trace is taken over colour structures, and the phase space
integration ${\rm d}\phi_n$ is schematically only referring to the
radiated partons in order to avoid notational clutter. The cross
section density operator for producing $n+1$ partons at a scale $q$
can iteratively be built up as
\begin{equation}
  {\mathbf A}_{n+1}(q) = \int_q^Q \frac{{\rm d}k}{k} {\rm P}e^{-\int_q^k \frac{{\rm d}k'}{k'}{\mathbf \Gamma}_n(k')}
  {\mathbf D}_n(k){\mathbf A}_{n}(k){\mathbf D}^\dagger_n(k)
  \overline{\rm P}e^{-\int_q^k \frac{{\rm d}k'}{k'}{\mathbf \Gamma}^\dagger_n(k')} \ .
\end{equation}
Several choices of evolution variables are possible. ${\mathbf D}_n$
describes the emission of the $n+1$st gluon (subject to the ordering
variable, {\it e.g.} energy) for which the colour structures 'grow'
into a larger colour space, and ${\mathbf \Gamma}_n$ encodes one-loop
virtual exchanges, which mix different colour structures into each
other \cite{Platzer:2013fha,Platzer:2020lbr}. These exchanges are
taken into account to all orders by means of the ordered
exponential. In practice we resort to the colour flow basis in colour
space, which allows for a convenient organization of an expansion
around the large-$N_c$ limit
\cite{Platzer:2013fha,AngelesMartinez:2018cfz}, including in principle
arbitrary higher orders in $1/N_c$. The process of gluon emission and
exchange in the colour flow basis is now known to two emissions, and
two loops, respectively \cite{Platzer:2020lbr}.  The CVolver library
\cite{Platzer:2013fha,DeAngelis:2020rvq} has been developed to solve
amplitude evolution equations in a new kind of parton shower
algorithm, and we have recently provided first results for jet vetoes
in $Z$ and Higgs decays, with further applications in progress. We
illustrate some results and the basic principles in
Fig.~\ref{fig:ampevolution}.
The work on amplitude evolution has also brought up the question to
what extent the mixing of colour structures through soft gluon
exchanges can be related to colour reconnection models. In fact, in
\cite{Gieseke:2018gff}, we have demonstrated that this might be the
case and the kinematic dependence, and the mechanism of Baryon
production through colour reconnection introduced in
\cite{Gieseke:2017clv} have a one-to-one correspondence with the
picture of soft gluon evolution. How we could include hadronization
effects in the amplitude evolution paradigm has up until now remained
elusive, and will be addressed in the remainder of this contribution.

\section{Constructing amplitude evolution}
\label{sec:constructingamps}

The cross section and evolution equations we have been advocating
above are relevant to jet cross sections where we can democratically
sum over colour configurations. This also implies that infrared
divergences, for infrared safe observables, will always cancel subject
to the cyclicity of the trace (see \cite{Forshaw:2021mtj} how the
cyclicity and multi-parton colour correlations can be used to extract
subleading-$N_c$ radiation patterns). We will now generalize this
picture by addressing exclusive final states, which is the target of
event generators and detailed predictions at the hadron level. Since
now projections of partonic systems onto certain hadronic systems come
into play, the most general from of the cross section should be
considered to be
\begin{equation}
  {\rm d}\sigma_m = \sum_n \int \alpha_0^n {\rm Tr}\left[{\mathbf M}_n {\mathbf
      U}_{nm}\right] {\rm d}\phi_n {\rm d}\phi_m \ .
\end{equation}
In this case ${\mathbf U}_{nm}$ will be the operator which projects
onto the observed hadronic systems (labelled by $m$) and contains the
observable function, and ${\mathbf M}_n$ is the cross section density
operator before any infrared and ultraviolet divergences have been
removed. In this expression, we integrate over possible partonic
momentum configurations ${\rm d}\phi_n$ and sum over partonic
multiplicities $n$. In case of a jet cross section, say, ${\mathbf
  U}_{nm}$ would be proportional to the unit operator in colour space
and would relate partonic to hadronic final states directly assuming
that hadronization would be a small correction. In the analysis below
we shall work with the effective measurement on parton level, thus
introducing
\begin{equation}
  {\mathbf U}_n = \sum_m \int
  {\mathbf U}_{nm}
  {\rm d}\phi_m \quad \text{and} \quad \sigma = \sum_m \int {\rm d}\sigma_m = \sum_n \int
  \alpha_0^n {\rm Tr}\left[{\mathbf M}_n {\mathbf
      U}_{n}\right] {\rm d}\phi_n \ .
\end{equation}
In general ${\mathbf U}_{n}$ defines an infrared unsafe cross section,
as implied by the hadronic final state we want to probe to formulate
how hadronization can enter the amplitude level evolution paradigm. In
\cite{Platzer:2022jny} we have detailed how a redefinition of the
measurement operator ${\mathbf U}_{n}$ in terms of a finite, though
possibly non-perturbative, measurement operator ${\mathbf S}_{n}$,
along with the standard redefinition of the bare coupling $\alpha_0$
in terms of the renormalized $\alpha_S(\mu_R)$ provides both UV and IR
subtractions for ${\mathbf M}_n$. A further redefinition of the
density operator ${\mathbf M}_n$ in terms of a then finite density
operator ${\mathbf A}_n$ provides an expression of the cross section
in terms of resummed and finite quantities. The cross section is in
particular renormalization group invariant both with respect to the UV
renormalization scales $\mu_R$ and the resolution scales $\vec{\mu}_s$
which separate out unresolved radiation. In an event generator, there
would be one scale $\mu_S$ which is the parton shower infrared cutoff
at which hadronization takes over from parton shower evolution, but
this is a certain choice and needs to be looked at more general. If
the redefinitions have not been truncated at fixed order in
$\alpha_S(\mu_R)$, the resulting cross section is
\begin{equation}
  \sigma = \sum_n \int
  \alpha_S^n(\mu_R) {\rm Tr}\left[{\mathbf A}_n(\mu_R,\vec{\mu}_S) {\mathbf
      S}_{n}(\mu_R,\vec{\mu}_S)\right] {\rm d}\phi_n
\end{equation}
and thus is independent of any of the resolution scales. A residual
resolution scale dependence arises at fixed order, and can be related
to a tower of subleading logarithms in the evolution
\cite{Platzer:2022jny}. Evolution equations in all of the separation
scales are implied. Within our formalism, we have the advantage that
we can relate these infrared resolution criteria to the classes of
observables in question and thus judge how accurate our algorithm will
be able to generate resummed predictions. More importantly, the RGE
evolution of ${\mathbf A}_n$,
\begin{equation}
  \label{eq:evolution}
    \partial_S {\mathbf A}_n = {\mathbf \Gamma}_{n,S}{\mathbf A}_n + {\mathbf A}_n {\mathbf \Gamma}_{n,S}^\dagger -
  \sum_{s\ge 1} \alpha_S^s {\mathbf R}_{S,n}^{(s)} {\mathbf A}_{n-s} {\mathbf R}_{S,n}^{(s)\dagger}
\end{equation}
exactly resembles the parton branching at amplitude level algorithm we
have been introducing in the first part, where ${\mathbf R}^{(s)}_n$
now describes emissions of $s$ partons on top of $n$ already radiated
partons and does itself include virtual corrections, {\it e.g.} a
one-loop correction to emission of a single gluon. The subscript $S$
on the derivatives simply indicates the relation to the specific
component of $\vec{\mu}_S$ we have been differentiating to. The
evolution equation for a non-trivial ${\mathbf S}_{n}$ (with a similar
notation as in Eq.~\ref{eq:evolution}, $[{\rm
    d}p_i]\tilde{\delta}(p_i)$ indicating the phase space of an
emitted parton $i$),
\begin{equation}
  \partial_S {\mathbf S}_n = -\tilde{\mathbf \Gamma}_{S,n}^\dagger {\mathbf S}_n - {\mathbf S}_n \tilde{\mathbf \Gamma}_{S,n}
  + \sum_{s\ge 1}\alpha_S^s
  \int \tilde{\mathbf R}_{S,n+s}^{(s)\dagger} {\mathbf S}_{n+s} \tilde{\mathbf R}_{S,n+s}^{(s)}
  \prod_{i=n+1}^{n+s}[{\rm d}p_i]\tilde{\delta}(p_i) \ ,
\end{equation}
can be seen as an evolution equation of a hadronization model: the
hard density matrix ${\mathbf A}_n$ evolves from a large scale $Q$
down to the infrared scales, and from lower to higher multiplicity,
while ${\mathbf S}_n$ evolves from small scales, and thus in principle
a non-perturbative initial condition, and from larger to smaller
multiplicities. Both objects also do evolve in opposite directions in
colour space dimensionality.  In the evolution of ${\mathbf S}_n$, we
recover in particular the mechanism of colour reconnection mentioned
in the first part. Other features depend on what model is used to
express the non-perturbative initial condition needed for ${\mathbf
  S}_{n}$. If, in particular, the a cluster-model like {\it Ansatz} is
used, then additional gluons can be absorbed in the evolution of
${\mathbf S}_n$ in the form of soft $q\bar{q}$ pairs, thus resembling
the cluster fission mechanism \cite{Bahr:2008pv}. The picture which
emerges puts the parton shower infrared cutoff (if there is a single
resolution scale) onto the same conceptual level as the factorization
scale for parton distribution functions, and indicates that the hard
parton shower evolution, and the evolution of the hadronization model
shall be matched at this scale. Any residual dependence on
$\vec{\mu}_S$ would then serve as a measure of uncertainty. In fact,
by perturbatively expanding out the evolution equation of ${\mathbf
  S}_n$, one can quantify how accurately a given observable, subject
to some infrared resolution, will be predicted by the algorithm, see
\cite{Platzer:2022jny} for more details. In a first step towards
precision \cite{Platzer:2022jny} also calculates all relevant
structure up to the second order and all anomalous dimensions and
emission operators are known to the two-loop level in the soft case
\cite{Platzer:2020lbr}, and we have made first steps to extent this
beyond the soft limit to systematically include collinear
contributions, as well \cite{Loschner:2021keu}.

\section{Conclusion}

In this contribution we have presented an overview of parton shower
development centred around the amplitude level evolution
paradigm. This framework is used both as a theoretical tool as well as
an algorithm in its own right, which is able to calculate observables
in a subleading-$N_c$ accurate way. We have focused on how such an
approach can be derived from a renormalization group point of view,
which gives rise to identifying how a hadronization model can be
included and constrained from amplitude evolution. Colour reconnection
models based on such an approach are directly reproduced, as are
features of cluster fission. We plan to further use this insight to
extent the hadronization models in Herwig, and to supplement the
CVolver approach by hadronization corrections.

\section*{Acknowledgements}

I am grateful to Jeff Forshaw, Jack Holguin, Maximilian L\"oschner,
Ines Ruffa and Malin Sj\"odahl for numerous discussions on this
subject and the fruitful collaboration on many related aspects.

\paragraph{Funding information}

This work was in part supported by the European Union’s Horizon 2020
research and innovation programme as part of the Marie
Sklodowska-Curie Innovative Training Network MCnetITN3 (grant
agreement no. 722104). I also thank the Erwin Schr\"odinger Institute
at Vienna for support through numerous Research-in-Teams programs
RIT2020, RIT0421 and RIT0521.

\bibliography{amplitude-evolution}

\nolinenumbers

\end{document}